\documentclass{article}

\usepackage[preprint]{neurips_2019}

\usepackage[utf8]{inputenc} 
\usepackage[T1]{fontenc}    
\usepackage{hyperref}       
\usepackage{url}            
\usepackage{booktabs}       
\usepackage{amsfonts}       
\usepackage{nicefrac}       
\usepackage{microtype}      
\usepackage{graphicx}
\usepackage{natbib}

\title{A blockchain-orchestrated Federated Learning architecture for healthcare consortia}

\author{
Jonathan Passerat-Palmbach \\
ConsenSys Health / Imperial College London \\
\texttt{j.passerat-palmbach@imperial.ac.uk} \\
\And
Tyler Farnan \\
ConsenSys Health / UC San Diego \\
\texttt{tfarnan@ucsd.edu}
\And
Robert Miller \\
ConsenSys Health \\
\texttt{robert.miller@consensys.net} \\
\And
Marielle S. Gross \\
Johns Hopkins Berman Institute of Bioethics \\
\texttt{mgross23@jhmi.edu} \\
\And
Heather Leigh Flannery \\
ConsenSys Health \\
\texttt{heather.flannery@consensys.net} \\
\And
Bill Gleim \\
ConsenSys Health \\
\texttt{bill.gleim@consensys.net} \\
}

\begin{document}

\maketitle

\begin{abstract}
We propose a novel architecture for federated learning within healthcare consortia. At the heart of the solution is a unique integration of privacy preserving technologies, built upon native enterprise blockchain components available in the Ethereum ecosystem. We show how the specific characteristics and challenges of healthcare consortia informed our design choices, notably the conception of a new Secure Aggregation protocol assembled with a protected hardware component and an encryption toolkit native to Ethereum. Our architecture also brings in a privacy preserving audit trail that logs events in the network without revealing identities.
\end{abstract}


\hypertarget{context-for-privacy-in-healthcare}{
\section{Privacy in healthcare and Federated Learning Consortia}\label{context-for-privacy-in-healthcare}}

The healthcare sector is uniquely positioned to leverage data for the purpose of creating value and improving human health. However, our ability to learn from health data is in tension with a unique set of ethical, legal, economic and technical challenges related to data privacy.
But traditional methods of mitigating health data privacy concerns, namely HIPAA and the use of deidentified data, have proven insufficient to protect individuals' interests. Realizing data's promise will require new tools, and we propose a novel federated learning architecture that is uniquely suited to these problems.


Large scale federated learning as depicted in the original series of
papers \cite{bonawitz_practical_2017, bonawitz_towards_2019} involves a great number of mobile devices
that take part in the training rounds. This introduces a series of
design constraints and challenges to enable learning to happen on these
resource-limited devices without disrupting the end-user's experience nor privacy.
We argue that among the challenges highlighted in \cite{bonawitz_towards_2019}, some elements are not relevant anymore when considering a
consortium context.

The most significant change lies in the data distribution. A consortium
will contain less parties than a public network of mobile devices, but
each of these participants will host a larger quantity of data. They
will also benefit from substantially more compute and storage resources,
as the devices involved in the training rounds will be server-grade
machines rather than
smartphones. Another discrepancy exposed by servers is that they can be
leveraged in a training round at any time of the day (contrary to
mobile devices which will be leveraged mostly when charging and connected to a home Wi-Fi network). Network
connectivity will also be more reliable, thus almost no participant will
drop out during a training round.

These elements informed our design choices, which as a
result differ from the blueprints established in \cite{bonawitz_towards_2019, lalitha_fully_2018}.  
The architecture we introduce in this paper draws its specificity and
novelty from the combination of four main features making it
particularly suitable for the healthcare consortia settings we are considering: \textit{1)} Data owners can define fine-grained access policy to their data,
  restricting access to certain members of the
  consortium only (Section \ref{fine-grained-data-permissioning}), \textit{2)} An alternative implementation of Secure Aggregation based on AMD SEV
  in memory encryption, which introduces a trusted third-party more
  suitable than a potentially brittle Multi-Party Computation (MPC) with
  a limited number of non-anonymous participants (Section \ref{amd-sev-based-secure-aggregator}), \textit{3)} Peer-to-peer transit encryption of updated weights between the
  workers and the Secure Aggregator leveraging cutting edge Ethereum
  technology \cite{alabi_ethereum_2018} (Section \ref{peer-to-peer-in-transit-encryption-of-updated-weights}), \textit{4)} A privacy preserving audit trail logs actions undertaken within the network while
  keeping the actual list of participants to a training round hidden (Section \ref{privacy-preserving-audit-trail}).


\hypertarget{system-architecture}{%
\section{System architecture}\label{system}}

\begin{figure}[tb]
    \centering
    \includegraphics[keepaspectratio, scale=0.33]{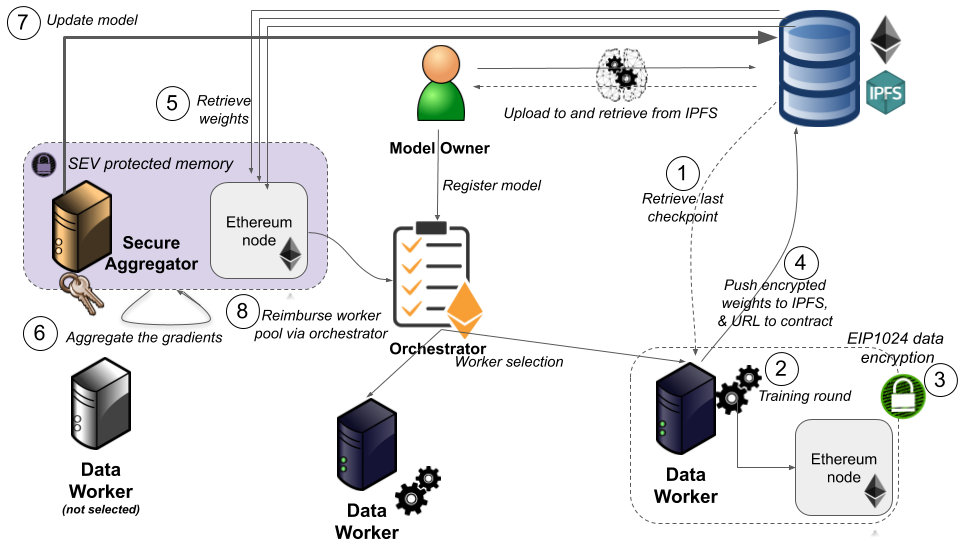}
    \caption{Global overview of a training round}
    \label{fig:architecture}
\end{figure}

Figure \ref{fig:architecture} gives an overview of the system architecture and and of a training round. All the parties in the network have access to their own Ethereum account that will identify them and enable them to interact with the rest of the network. Our solution exploits the Federated Learning building bricks provided by the PySyft library \cite{ryffel_generic_2018}.

\hypertarget{overview-architecture}{%
\subsection{Overview}\label{workflow}}

The \textit{Model Owner} (MO) initiates the process by uploading his model to a distributed storage infrastructure (IPFS, Swarm, \dots) available to all the members of the consortium. MO then registers the model and a training description to an \textit{Orchestrator} smart contract. This description will contain similar information to what a \textit{FL Plan} does in \cite{bonawitz_towards_2019}.

The \textit{Orchestrator} is embodied by one or potentially several smart contracts. In addition to storing the models, it maintains a list of \textit{Data Workers} (DW). A DW is a node in the network that possesses one or many datasets that can be of interest for MOs to improve their models. DWs are communicating by instantiating a web socket server as available in PySyft. DWs have the ability to specify a blacklist of model owners that shouldn't have access to some or all of their data. They do so by updating a mapping of Ethereum addresses (representing the blacklisted MOs) and dataset identifiers stored in the \textit{Orchestrator} smart contract.

In order to initiate a training round, the \textit{Orchestrator} first determines the full set of data workers that comply with the previous exclusion rules (a post training selection will further reduce the set as described in \ref{random-selection-of-contributions-to-aggregate}). The Orchestrator schedules a training task for the selected workers using private transactions. All the actions undertaken by the Orchestrator are logged on chain in a privacy preserving fashion as described in Section \ref{privacy-preserving-audit-trail}. Each selected worker now pulls the latest checkpoint of the model from the shared distributed storage. The worker performs a training step as defined in the corresponding training description stored in the Orchestrator. Once it has completed its training round, the worker encrypts the newly obtained set of weights for the \textit{Secure Aggregator} (SA).

SA is a special entity, a Virtual Machine or container, whose memory is encrypted from its host using the AMD SEV technology (Section \ref{amd-sev-based-secure-aggregator}). The Aggregator fetches all the encrypted weights from the decentralised storage, decrypts them within the safe realm of its encrypted memory and performs the aggregation. It finally uploads the new model checkpoint to the shared storage and updates the Orchestrator smart contract with new pointer. Multiple SA could be instantiated to accommodate a larger workload or introduce more decentralisation and reliability.

\hypertarget{fine-grained-data-permissioning}{%
\subsection{Role of Ethereum and Fine grained data
permissioning}\label{fine-grained-data-permissioning}}

In the context of federated learning consortia, it is critical to prevent legal, ethical, and competitive requirements from being compromised. We argue in this paper that the Ethereum \cite{wood_ethereum:_2014} blockchain can fulfill such a role and benefit an architecture targeting consortia in the healthcare industry in particular. The combination of a decentralised immutable ledger that can be updated programmatically in a highly trustable flavour makes Ethereum a very appealing choice to design modern decentralised and secure systems. Hyperledger Besu is an enterprise-grade Ethereum client. These tools enable consortia to implement cooperative standards for viewing, transacting, and communicating within in the network. For example, providing read-only access of an audit trail to an external third-party or restricting data exchange to a subset of network nodes for private communication.

On top of Besu’s native features, an Orchestrator smart contract controls the traffic in the network, keeping track of permissions set by each user regarding the level of data-sharing access with other members in the consortium. In addition to governing activity and authorizations, smart contracts can also support the exchange of native tokens within the Ethereum blockchain. Tokens have been proposed as a way of incentivizing network participants to perform actions; in our architecture they could be used to incentivize computations. However, they extend naturally as a fungible or non-fungible measure of value in an information exchange, and can serve as the backbone for incentivized data-sharing in a federated learning consortia. In our use case, smart contract(s) can record metrics throughout federated learning lifecycles, and specify the execution of remuneration policies chosen by the consortia.

\hypertarget{amd-sev-based-secure-aggregator}{%
\subsection{AMD SEV based Secure
Aggregator}\label{amd-sev-based-secure-aggregator}}

Architectures like \cite{bonawitz_towards_2019}
are designed for training rounds involving ``a few hundred devices''. Within the context of healthcare, federated learning consortia are mostly likely to be adopted by enterprises first. This will limit the number of members in a network, likely to be under 100. This number will shrink even more when
applying the data selection filters and account permissioning rules.

While medical applications have been shown to successfully leverage
federated learning with a lower number of workers (20-30) \cite{roy_braintorrent:_2019, crimi_multi-institutional_2019} compared to \cite{bonawitz_towards_2019}, this could break
the security properties of the Secure Aggregation scheme presented in
\cite{bonawitz_practical_2017}. Secure Aggregation \cite{bonawitz_towards_2019} is
performed via a MPC protocol built on top of
Shamir's Secret Sharing (SSS) scheme \cite{shamir_how_1979}. In our present
context, the number of active devices would be too low to provide
the same level of guarantees for the MPC. On top of that, members of a
consortium are highly likely to know each other and this would
dramatically increase the risks of compromising the security threshold
of the SSS scheme.

We introduce a better suited approach for such consortia based on protected hardware such as AMD's
Secure Encrypted Virtualization (SEV) memory encryption technology and Intel’s Software Guard
Extensions (SGX). \cite{mofrad_comparison_2018} provides a
concise comparison of these technologies, each of which generally cryptographically isolate software containers from the host system to
better protect confidential data.

Secure Aggregation as described in \cite{bonawitz_practical_2017} is secure
in the \emph{honest but curious} context. That is, if an attacker gains
access to a participant in the MPC, he will only find secret
shares from which he will not be able to reconstruct the private data.
Here, SEV provides equivalent guarantees of in-memory privacy from attackers
outside the VM/container it protects. As a result, we implement a
Secure Aggregator as a trusted third party running inside an AMD SEV
protected VM. At the end of a training round, workers communicate their encrypted computed weights to the Secure Aggregator who performs the aggregation then uploads a non-encrypted updated model back to the shared storage.

\hypertarget{peer-to-peer-in-transit-encryption-of-updated-weights}{%
\subsection{Peer-to-peer in transit encryption of updated
weights}\label{peer-to-peer-in-transit-encryption-of-updated-weights}}


Weight encryption while in transit between worker and aggregator is
however extremely important since it contains the private data used for
this training step that should remain private to the worker. We assume
here that the members of the consortium will adopt honest but curious behaviour, and would have a strong incentive to
eavesdrop on the network to learn of the raw updates shared by other
members containing valuable information on their private datasets.

At the time of writing, encryption over the wire is not available to protect the traffic between two workers in PySyft. While this feature is under active
development, our system will leverage Ethereum Improvement
Proposal (EIP) 1024 \cite{alabi_ethereum_2018} that offers peer-to-peer transit encryption of
arbitrary data between two Ethereum accounts. Using EIP-1024, we can encrypt the new weights calculated by each worker before they are being sent back to the Secure Aggregator.

\hypertarget{security-requirements}{%
\section{Security requirements}\label{security-requirements}}

Secure aggregation brings a first level of privacy by hiding the raw
updates to a model from a given data owner. However, the
identity of the data owners who took part in a training round could
allow an attacker to extract meaningful information from the aggregated
weights. We describe two counter-measure to this issue in this section.


\hypertarget{random-selection-of-contributions-to-aggregate}{%
\subsection{Random selection of contributions to
aggregate}\label{random-selection-of-contributions-to-aggregate}}



For a given federated training round, all chosen workers will compute gradient updates, but the Secure Aggregator will randomly select a subset of weights for aggregation. After each training round, the Secure Aggregator will forget the unpicked weight updates and the workers will not be notified whether or not their updated weights were selected for aggregation. 

In the most critical case, a malicious worker or a cartel of workers could communicate their weight updates for a given training round via a third-party channel. This would allow the model owner to reverse engineer a round update and extract the exact contribution of each worker as highlighted in \cite{melis_exploiting_2018}. By selecting updates for aggregation, such malicious intent to expose another member’s data contribution can be prevented. Note that while this privacy preserving mechanism is distinct from Differential Privacy (DP) \cite{papernot_semi-supervised_2016, abadi_deep_2016}, they can be combined to provide even more enhanced privacy guarantees.

\hypertarget{privacy-preserving-audit-trail}{%
\subsection{Privacy preserving audit
trail}\label{privacy-preserving-audit-trail}}

As described in the global overview of the architecture, a residual benefit from our Ethereum-based system is the immutable audit trail stored
on chain. The audit trail gathers events related to the learning
process. As part of our privacy in depth proposal, we must not leak sensitive information that may compromise the
privacy guarantees of the federated learning process.

The Secure Aggregator (SA) will spearhead this non-identifiable update
protocol. First, for each data worker taking part in the round the SA
will generate a new unique random nonce for the current round. Each
worker already has an established secure communication channel with SA
given EIP 1024 transit encryption. Data between SA
and a worker are encrypted using a symmetric Diffie-Helmann (DH) secret
derived from the worker's public key and SA's private key (or
conversely). SA will apply Ethereum's \texttt{keccak256} function to the
concatenation of this shared DH secret and the nonce it just generated, sending the resulting hash encrypted via EIP 1024 to the
corresponding worker. The hash is published on chain and will serve
as an anonymous identifier for the corresponding data worker during a training round.

The only attack to uniquely identify a data worker in the audit trail is to
know the DH secret that has been concatenated to the random nonce. Thus,
as long as the worker's and SA's respective private keys remain private,
they are the only two entities in the system able to reidentify the
worker's corresponding entries in the audit trail.



\hypertarget{conclusion}{%
\section{Conclusion}
}\label{conclusion}

This work introduced an architecture for federated learning in healthcare consortia built on the Ethereum blockchain. The architecture leverages Ethereum and its ecosystem (e.g., EIP 1024 encryption and the Besu enterprise client) to provide a coherent design reflecting the specific challenges in healthcare consortia. The novel architecture relies on a series of four "privacy in depth" blocks that provide a unique combination of features: fine-grained data access policies, a new Secure Aggregation agent running in hardware-protected processes, weight encryption via EIP 1024 and a privacy preserving audit trail of the events in a training round.

\bibliographystyle{abbrvnat}
\newpage

\end{document}